\documentclass[epsf]{aa}
\usepackage{epsf}

\begin{document}

\newcommand{\bd}{\begin{displaymath}}
\newcommand{\ed}{\end{displaymath}}
\newcommand{\be}{\begin{equation}}
\newcommand{\ee}{\end{equation}}

\title{Correlated X-ray and optical variability in KV UMa}

\author{H.C.\ Spruit\inst{1},\and G.\ Kanbach\inst{2}}

\offprints{H.C. Spruit(henk@mpa-garching.mpg.de)}

\institute{Max-Planck-Institut f\"ur Astrophysik,
           Postfach 1317, 85741 
           Garching bei M\"unchen, Germany \\
\and
           Max-Planck-Institut f\"ur Extraterrestrische Physik,
           Postfach 1312, 85741 
           Garching bei M\"unchen, Germany}
           
\date{Received 28-03-02; accepted 21-05-02}

\markboth
{Correlated X-ray and optical variability in KV UMa}
{X-ray and optical variability in KV Uma}

\abstract{
We present coordinated X-ray and optical observations of the X-ray transient KV UMa
(= XTE J1118+480) at high time resolution. The optical variation associated with spikes in the
X-rays consist of a dip preceding the spike by  0-5 s, and a peak following it. The
cross-correlation function between X-rays and optical has the same shape. It rises sharply,
within 30ms of the peak of the X-rays. The shape of the cross correlation shows significant
intrinsic variability on time scales as short as 30s. Analyzing this variability with principal
component analysis (calibrated with simulated data), we find two statistically independent
components. The first is similar to the  average cross-correlation. The second also consists of a
dip followed by a peak, but on a 3x shorter time scale. These properties of the optical light,
together with the high optical  brightness of the transient, are not easily explained by
reprocessing of the X-rays. 
}
\maketitle

\section{Introduction}

The transient X-ray source XTE J1118+480 (=KV UMa) appeared in January 2000 
(Remillard et al. 2000) and (with an
interruption in February) was active as an X-ray source till end of July.  The source appears to be
a relatively nearby (2 kpc) binary with an orbital period of 4 hrs, and a K-star companion. Its
mass function of 6 M$_\odot$ makes the primary a solid black hole candidate, (Mc Clintock et al.
2001a, Wagner et al. 2001). The X-ray outburst was somewhat unusual compared with
`standard' transients such as X-ray Nova Muscae. The X-ray flux was relatively constant during
the main outburst episode (March--July). The X-ray luminosity (1-100 keV) was only about
$10^{36}$ erg/s, and the source was always found in a X-ray hard state (most of the X-ray
energy flux appearing around 100keV). Since hard state spectra are the rule during the decay
phase of X-ray transients, this suggests that the accretion rate of the system was significantly
below the Eddington value. 

Due to the low galactic absorption at the position of the source, its spectral energy distribution
(SED) could be determined from the IR to X-rays, including the crucial EUV part of the spectrum
(Hynes et al. 2000, McClintock et al. 2000b), a unique result for X-ray transients. These
observations showed that the source had an unusually high optical-to-X-ray flux ratio compared
with other transients. The SED appears to peak at 100 eV, and the inferred UV-to-EUV flux of
$10^{36}$ erg/s was about the same as
the X-ray flux. The optical-to-X-ray flux ratio was about 0.1. High optical brightness during a
hard X-ray state has been seen before in the unusual long-term variable source GX 339-4 (Motch
et al. 1983).

The long duration of this transient, its location above the galactic plane, and its relatively close
distance made it an ideal target for simultaneous X-ray and optical observations. Though such
coordinated observations have been attempted regularly, successful observations at high
(sub-second) time resolution have been reported previously in only two cases. Motch et al.
(1981) reported correlated fast optical and X-ray variability from 96s of observations of GX
339-4, when it was in one of its rapidly variable and optically bright states. The X-ray-optical
cross-correlation of these data was reported to show a {\it negative} correlation
{\it preceding} the X-rays by a few seconds. Hynes et al. (2000) reported simultaneous HST and
XTE observations of the black hole transient GRO J1655-40. The cross-correlation showed, in
contrast with the case of 339-4, a positive correlation peak {\it following} the X-rays, interpreted
by the authors as evidence of reprocessing of  X-rays by the accretion disk. 

The optical emission of black hole candidates in outburst thus appears puzzling, both in its
unpredictable amplitude, and its contradictory correlations with the X-rays. Theoretically, the
optical light is interesting because it may provide a way to probe the structure of the accretion
flow, if the light is caused by reprocessing of X-rays from the central object (e.g. Arons \& King
1993). Evidence for such reprocessing of X-rays by an accretion disk has been found in type-1
X-ray bursts from accreting neutron stars (Matsuoka et al. 1984, Turner et al. 1985, van Paradijs
et al. 1990, Kong et al. 2000).

On the other hand,
lower limits to the optical brightness temperatures inferred from the short optical variability in GX
339-4 led Fabian et al. (1982) to conclude that a more likely mechanism for the optical radiation in
the Motch et al. data is thermal cyclosynchrotron emission in a strong magnetic field. We reached
a similar conclusion in a first description and interpretation of the results reported here (Kanbach
et al. 2001). If this interpretation is correct, the regions contributing most to the optical-UV light
would be much closer to the black hole than in a reprocessing model, at least in objects like KV
UMa and GX 339-4, and optical light of black hole transients would potentially be a powerful new
diagnostic of the inner accretion flow. 

In Kanbach et al. (2001) we found that the average X-ray/optical cross correlation function
consists of a peak starting at zero lag, preceded by a `precognition dip'. It thus combines
both of the effects seen previously by Motch et al. and Hynes et al. This raises the question how
these two components of the cross correlation are related, in particular since the dip at negative
lag is not very easily explained in models proposed so far. In this paper we address this question
by a  more detailed statistical analysis of our observations of KV UMa. The theoretical
interpretation  will be developed further in a future paper.

\section{Observations}
X-ray and optical observations were collected on the nights of 4,5,6 and 7 July 2001. X-ray
observations were made by the Rossi X-ray Timing Explorer, for about one orbit on each of
these nights. The total duration of useful simultaneous observations was 2.5 hours.  
The optical observations were done with the OPTIMA photometer (Straubmeier et al. 2001)
attached to the 1.3 m telescope on Mt.\ Skinakas, Crete (http://observatory.physics.uoc.gr/),
which is operated jointly by the Foundation for Research and Technology Hellas (FORTH) and
the Max-Planck-Institut f\"ur Extraterrestrische Physik. The photometer consists of a cluster of 
fiber-coupled avalanche diodes. The diodes are sensitive from 450 to 950 nm at a mean quantum 
efficiency of 50\%. Individual photon arrival times are recorded to an accuracy of 2$\mu$s 
with a GPS-based clock, at a maximum count rate of 200\,000/s.
On several of the nights, the telescope was facing into a fairly strong wind, causing modulations
of the signal at frequencies around 9 and 40 Hz due to telescope vibration. Most of the results
reported here are based on the correlation between optical and X-ray signals, for which this
artefact has little consequence. In the cross-correlation functions it appears as a source of noise
but has no  systematic effects.

The average total X-ray count rate was 500 s$^{-1}$ (3 detectors), typical for the brightness of the source during most of its outburst. The optical brightness was around V=13, resulting in a
count rate of about 24000 $s^{-1}$. While the X-ray flux was highly variable, with amplitudes
typical of the hard states of black hole transients or Cyg X-1, the rms variability of the optical flux
was only about 10\% on time scales of 100ms--30s. 

\begin{figure}
\mbox{}\hfill\epsfxsize1.0\hsize\epsfbox{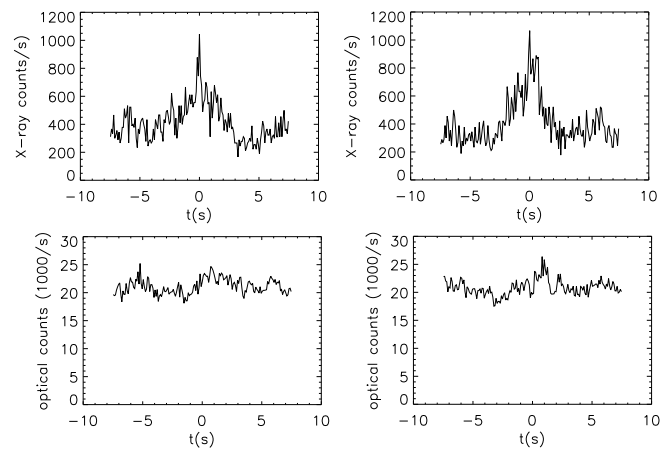}\hfill\mbox{}
\caption{\label{xocounts} Two examples of the X-ray and optical light curves, 
around the time of peaks in the X-rays.
}
\end{figure}

\section{Analysis}
An example of the X-ray and optical light curves is shown in Fig. \ref{xocounts}. The data show
the original count rates, without background or baseline subtractions. Peaks in the X-rays with 
durations of a few seconds as seen in this example were fairly typical. Visual inspection of such 
events suggests that the corresponding optical signal consists of a dip, preceding the X-rays by 
several seconds, followed by a positive response peaking a few seconds after the X-rays. 

In order to quantify this impression in a more objective way, an automatic peak-finding algorithm
was devised to collect and superimpose these events. The X-ray time series was compared with a
box-car smoothed version of itself (with a width of 30s). Time intervals where the actual signal
was more than twice the smoothed signal were identified as peaks. The 100 events found in this
way were aligned in time on the maxima of the X-ray peaks. 

\begin{figure}
\mbox{}\hfill\epsfxsize1.0\hsize\epsfbox{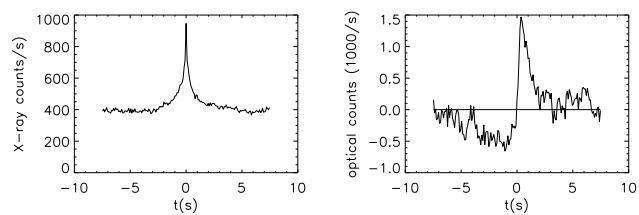}\hfill\mbox{}
\caption{\label{avglc} Average X-ray and optical light curves of 100 peak-aligned events like
the ones shown in Fig.\ \ref{xocounts}.
}
\end{figure}

\begin{figure*}[t]
\mbox{}\hfill\epsfysize0.3\hsize\epsfbox{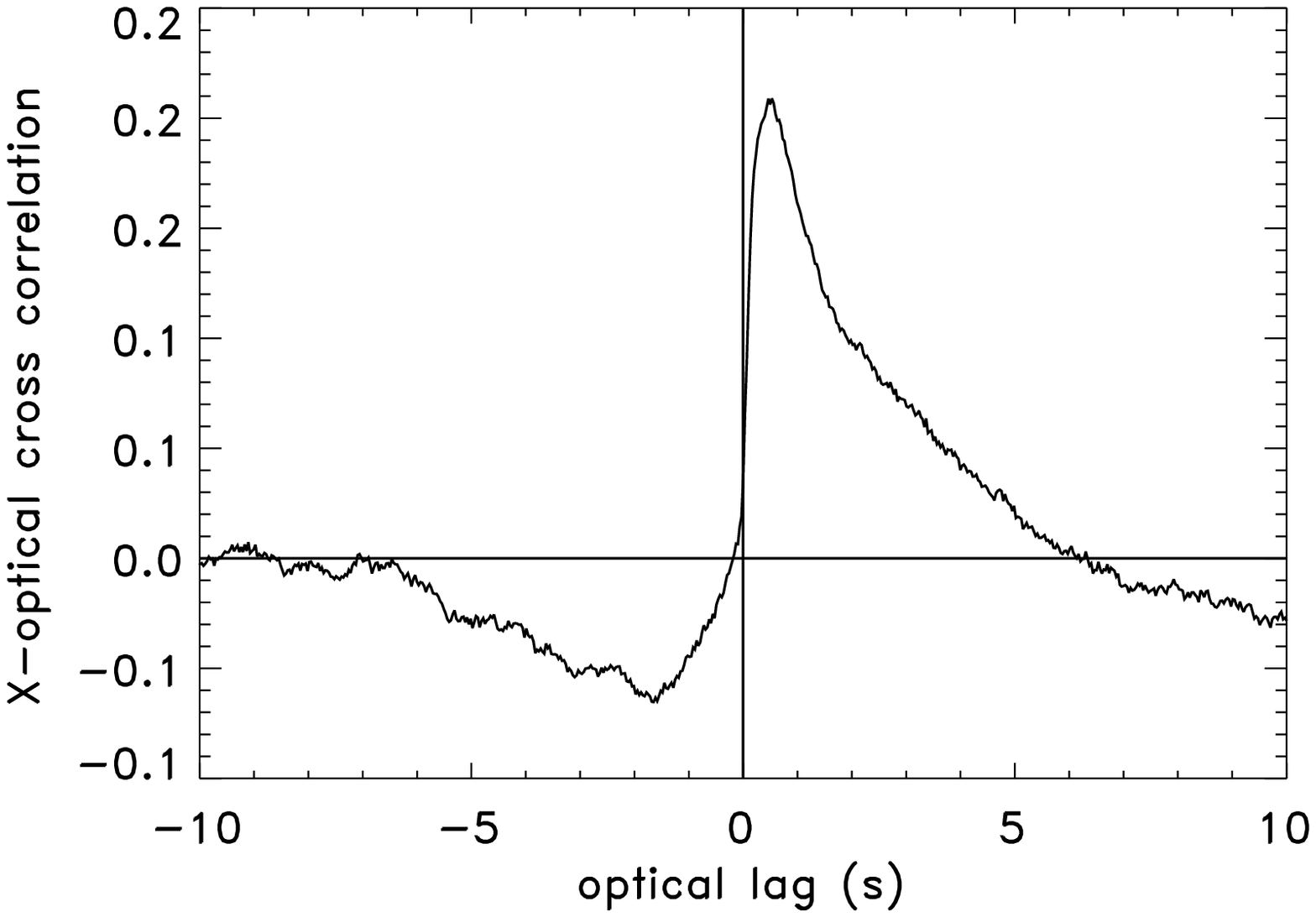}\hfill\epsfysize0.3\hsize\epsfbox{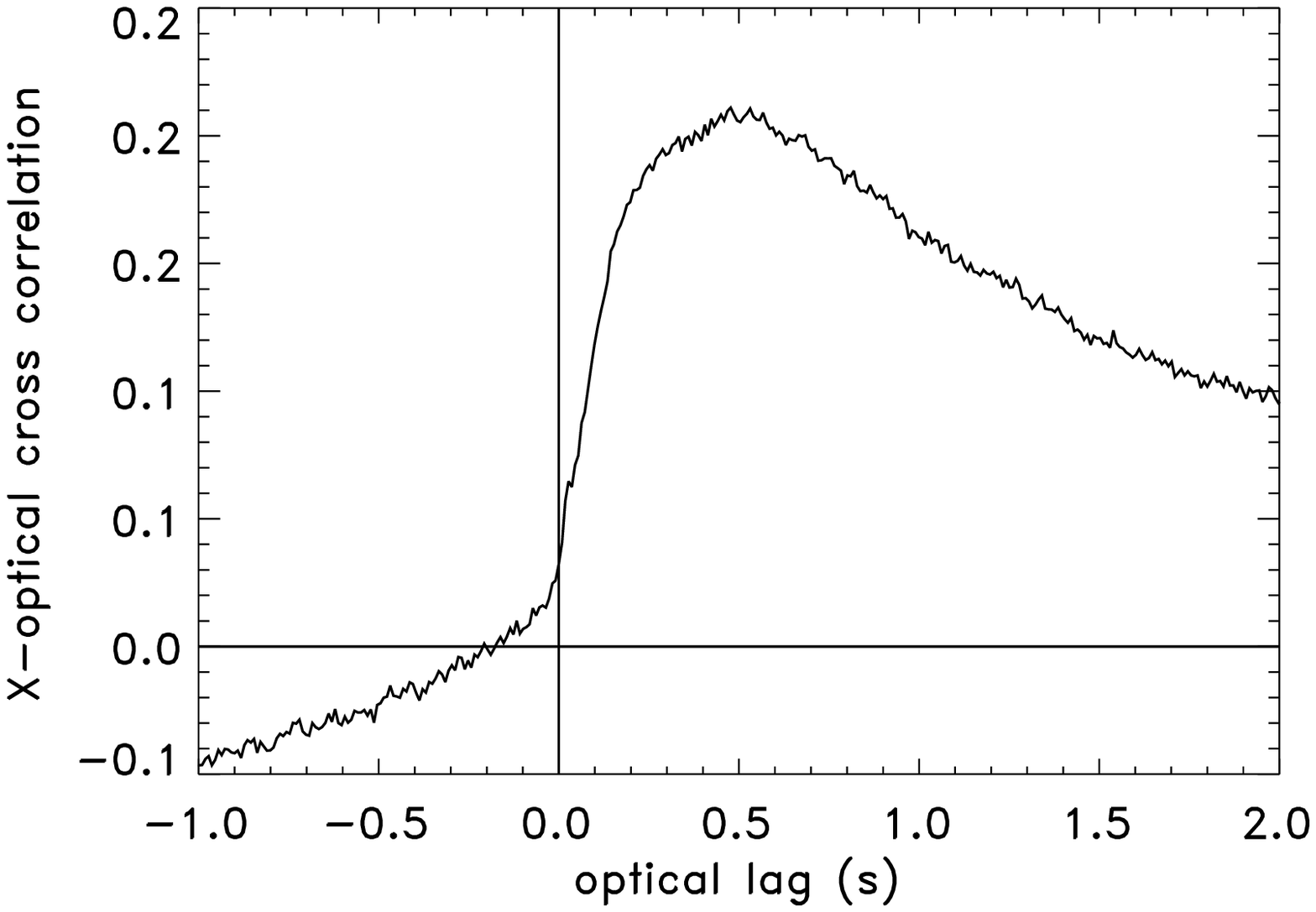}\hfill\mbox{}
\caption{\label{avgcxo} Average X-ray/optical cross correlation from 297 sample of the time
series. Right panel shows the cross correlation on an expanded scale; note the sharp rise at zero
lag.
}
\end{figure*}

The average X-ray and optical light curves of the peak events found in this way is shown in
Fig.\  \ref{avglc}.  This average confirms the subjective impression given by Fig.\
\ref{xocounts}. The dip in the optical preceding the X-rays appears to be a systematic effect,
though its amplitude varies between events. 

\begin{figure*}[htb]
\mbox{}\epsfysize0.7\hsize\epsfbox{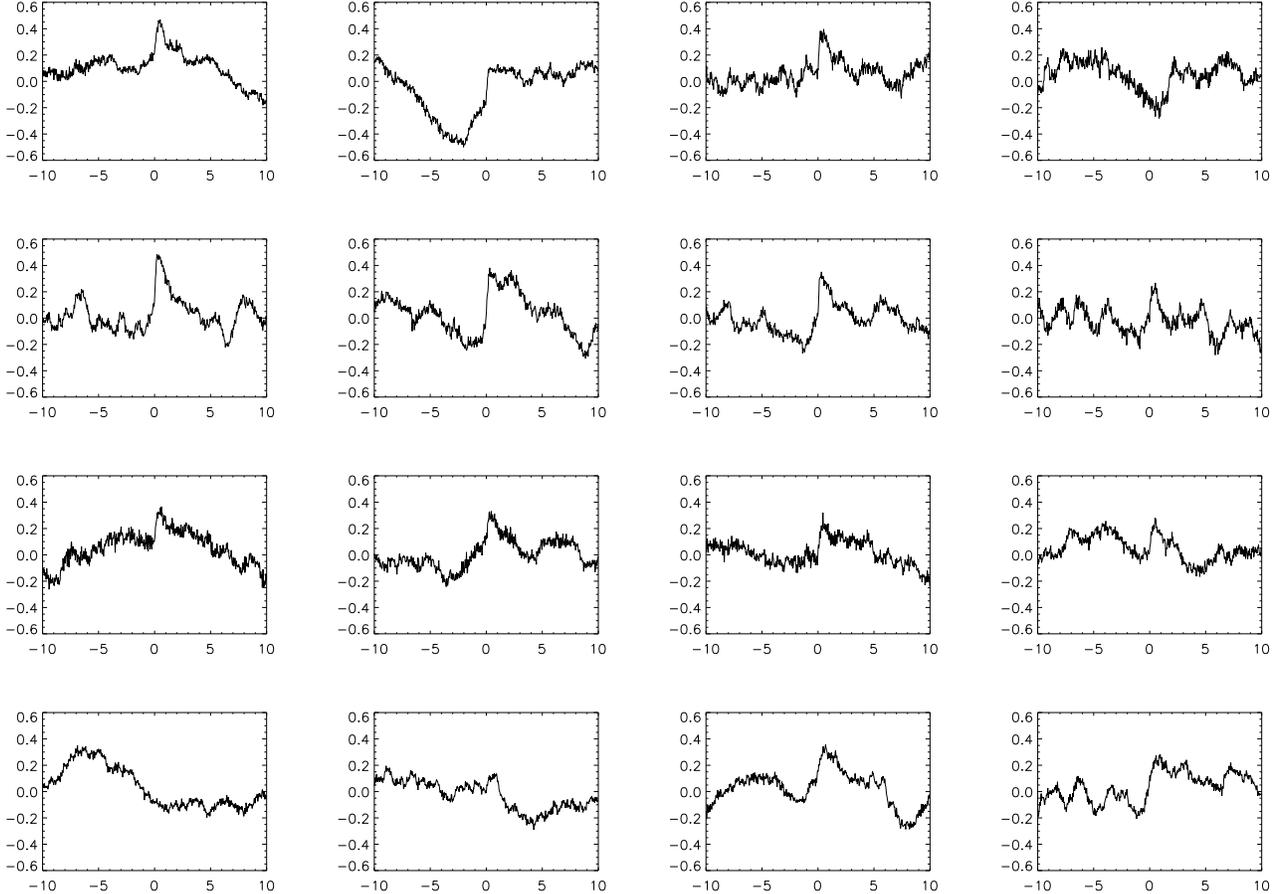}\hfill\mbox{}
\caption{\label{varc} 16 randomly selected samples from the set of 297 X-ray/optical cross
correlation functions. 
}
\end{figure*}

\subsection{Cross-correlation}
The superposition of events shown in Fig.\ \ref{avglc} is limited in its time resolution by the 
width of the X-ray peaks, of the order of a second. A view at much higher time resolution, at the
expense of a somewhat less direct interpretation, is given by the cross correlation between the
data. The cross-correlation of the X-ray time series $S_{\rm X}$ and the optical series $S_{\rm
O}$,  $C_{XO}(t)= S_{\rm X}*S_{\rm O}$, where a $*$ denotes the convolution, was computed 
for 297 independent segments of data of 25s length. The average of these cross-correlations is 
shown in Fig.\  \ref{avgcxo}. 

The cross correlation contains a residual of the photon noise in the original time series. In Figs.\
\ref{avgcxo}, \ref{varc} this shows up as the weak high-frequency noise component. Its
amplitude
is much lower, however, than the noise due to the intrinsic variability of the cross correlation, as
Fig.\ \ref{varc} shows. 

The cross-correlation appears to contain significant signal for lags (optical later than X-rays) from
about -5 to +5s. On these time scales, the correlation shows the same properties as the direct
comparison of the signals shown in Fig. \ref{avglc}: a dip followed by a peak of about the
same amplitude. 

As the right panel of Fig.\ \ref{avgcxo} shows, the change from the dip to the peak takes place
quite close to zero lag. The slope of the correlation increases by a factor 10 within about 30 ms
at  $t=0$, and 50\% of the peak is reached at $t=100$ ms. 

\subsection{Dip and peak}

A peak starting at $t=0$ is understandable in a reprocessing model for the optical light, and its
duration is also of the order expected for reprocessing of X-rays by an accretion disk. Evidence
for such reprocessing has been seen in X-ray bursters (Matsuoka et al. 1984, van Paradijs et al.
1990, Kong et al. 2000). It has also been advanced to explain the optical response in some black
hole transients (O'Brien \& Horne 2000, Hynes et al. 1998).  A delayed optical response is also a
natural result in the outflow model of Kanbach et al. (2001).  In this model, the unusually large
optical flux of KV UMa does not come from the accretion disk but  instead is produced by
synchrotron emission in a magnetic outflow. The dip in the optical light preceding the X-rays is
not very
natural in either of these explanations, however. This raises the question whether the dip should
be regarded as an independent phenomenon, or as a part of the correlation function associated
directly with the peak. 

The question whether the dip represents an independent physical process from that producing
the peak can be answered to some extent by investigating the variability of the dip and the peak
between different samples of the cross correlation. If their amplitudes turn out to be statistically
uncorrelated, it is unlikely that they are due to the same process. Significant variability is in fact
present in both the dip and the peak. This is shown in Fig.\ \ref{varc}, which presents a random
subset of 16 of the 300 correlations. From this figure it is also clear, however, that a large part of
the variability of the cross correlation appears to be independent of both the dip and the peak.
In many cases, this additional component masks the dip and/or the peak. From simple inspection
of figures like these, it is hard to determine if this additional variability is just some form of
intrinsic random variability in the source, or whether it contains additional systematic variations.

\section{Principal components of the cross correlation}
A more quantitative way of investigating the variability of a signal like the cross correlations in
Fig.\  \ref{varc} is by a principal component analysis (PCA, e.g. Kendall 1980, not to be 
confused here with the PCA detectors on board of XTE). If the signal is
variable, and a sufficiently large set of samples of it available, this method can decompose the
samples into components whose amplitudes vary statistically independently through the set. The
method has been used  in astronomy, among other applications, for the classification of galaxy
spectra from large surveys (e.g.\ Folkes et al. 1999), AGN spectra (Boroson 2002), Chepheid
light curves (Kanbur et al. 2002), and velocity distributions in molecular clouds (Brunt \& Kerton
2002). The components found in this way can often be interpreted more easily in terms of
variations in physically quantities than the original (large) datasets. 

In practice, significant care has to be excercised in interpretating the principal components. In the
following we do this by comparison with forward modeling. First we apply PCA to our set of
297 samples of the cross correlation and interpret a few of the most significant components. We 
then verify the stability and significance of these components by constructing sets of synthetic 
cross correlation functions, and processing them in the same way. These synthetic samples 
consist of superpositions of a few input components whose shapes are guessed, with statistical 
properties similar to those of the real set of cross correlations. The input components and the 
statistics of their amplitudes are adjusted to find the minimal number of `true' components, and 
their shapes, that are needed to reproduce the pattern of components found by the PCA of the 
real data set.

\subsection{Method}

For our application, the PCA is used as follows. Let the X-O cross correlation function of sample
$j$ be denoted by ${\bf a}^j(t)$. {Let $m$ be the number of available samples of the
correlation function ($m=297$ in the present application), discretize the time-delay axis $t$  in $n$
bins $t_i$  ($-12.5<t_i<12.5$s, $n=150$ in our case), and} write $a_i^j={\bf a}^j(t_i)$. We treat the $n$ time
bins as the variables whose statistical properties are to be found by PCA. Since neighboring time
bins are highly correlated, the number of degrees of freedom is much less than $n$. The degree
of correlation of the time bins with each other contains information on the actual number of
degrees of freedom in the signal. We measure the correlation between time bins $i$ and $k$ by
the covariance matrix $C_{ik}$:
\be C_{ik}=\sum_j (a^j_i-\bar a_i)(a^j_k-\bar a_k),\ee
where $\bar {\bf a}=\sum_j {\bf a}^j/m$ is the average of ${\bf a}$ in the set. The principal
components of ${\bf a}$ are then conventionally defined as those linear combinations of the
columns of $a_i^j$ that diagonalize the covariance matrix. [In many PCA applications, the
correlation matrix is used instead of the covariance matrix. This is important if the variables
measured are different physical quantities. For our case, this makes practically no difference,
since the variables $a_i$ are the discretized values of a single continuous function.] If the
coefficients of these linear combinations are denoted by the matrix $c_{ik}$, the principal
components $p_k$ are given by
\be p_k^j=\sum_i c_{ki}a^j_i, \ee
and the samples themselves are given, in terms of the principal components, by the inverse of
$c$,
\be a^j_i=\sum_k c^{-1}_{ik}p_k^j.\ee

These components are somewhat abstract, since a linear combination of different time bins is not
a physically very meaningful quantity in our application. A more useful representation of the
principal components is found by noting that the quantity 
\be a_{ki}^j=c^{-1}_{ik}p_k^j \ee
describes the amplitude of principal component $k$ in the $i$-th time bin of the $j$-th sample. The
average of this amplitude over the set of samples,
\be P_{ki}\equiv\sum_ja_{ki}^j/m,\ee
describes the principal components in the `time domain', i.e. as a decomposition of our average
cross correlation function. We use this representation for the rest of our analysis. The average
of the cross correlation function over the set, $\bar {\bf a}$, is the sum of these principal
components:
\be \bar a_i=\sum_k P_{ki}.\ee

\begin{figure}
\mbox{}\hfill\epsfysize1.25\hsize\epsfbox{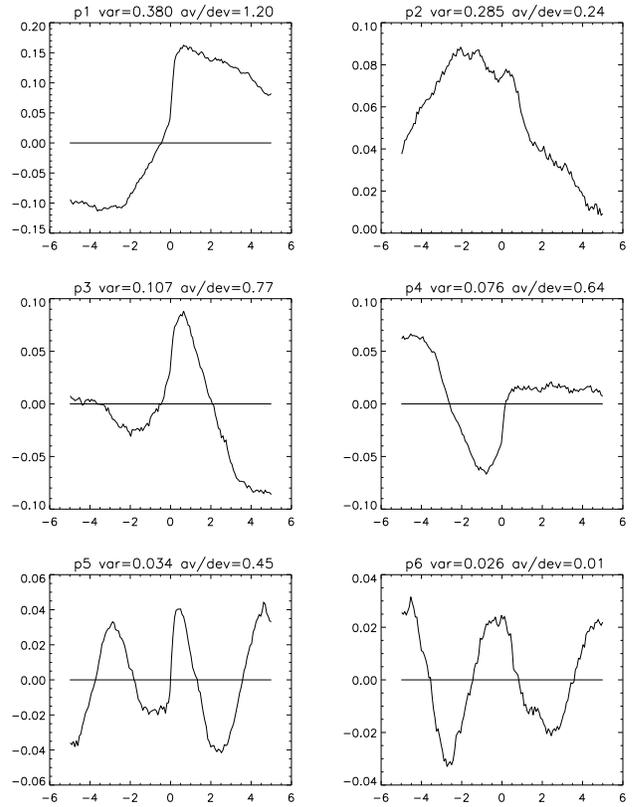}\hfill\mbox{}
\caption{\label{allsix} The first six principal components from a PCA of the set of 297 samples of 
the X-ray/optical cross correlation. Numbers above the panels show the fraction of the variance 
in the cross correlation accounted for by the components, followed by ratio of the average to the
deviation of the component's amplitude.
}
\end{figure}

\subsection{Results}
The covariance matrix is singular if the number of time bins $n$ is larger than the number of
samples $m$. This limits the time resolution that can be used for the discretization of our cross
correlation functions to 300 points\footnote{This limitation can be circumvented by 
data-compression of the samples. A convenient compression is obtained by applying a 
wavelet-transform to the samples. PCA is then applied on the first $n_0\le m$ wavelet
coefficients, and the resulting principal components transformed back to the time domain by the
inverse wavelet transform. For $n_0=n$ this process is loss-free and yields identical results}.  The
results depend somewhat on the range in $t$ used for the analysis. Most of the signal of interest
appears between $t=\pm 5$s. Using this range, and bins of 30ms yields the principal components
shown in  Fig.\ \ref{allsix}. 

The amplitude $A$ of components varies both in absolute value and sign. For the first
component we find that the average is larger than the variation, but for the higher components
the average is generally small compared to the variation.  Instead of showing the components at
their
average amplitude $\bar A$, we have therefore plotted them at a scale that takes into account
both the average and the variation.  The value used for this, somewhat arbitrarily,  is $\tilde
A=(\bar A^2+\sigma_A^2)^{1/2}$, where $\sigma_A$ is the rms deviation of $A$. 

The first two components together account for 67\% of the variance between the samples of the
cross correlation function, if a time-lag interval $(-\Delta t <t<\Delta t)$ with $\Delta t=5$ s is
used. The first component appears quite similar to the average cross correlation. The second
component peaks around the `precognition dip', though it has the opposite sign and
extends somewhat to positive lag as well. If the time interval used is shorter, the first two
components account for a larger fraction of the variation, and less for a longer interval. For
$\Delta t=2 s$ the first two components account for 86\% of the variance, for $\Delta t=12$ s
only 44\%. This is understandable if the cross correlation consists of a signal of relatively stable
shape concentrated around $t=0$, plus an unsystematically varying component that is spread out
over a wider range in the time delay coordinate $t$. 

\begin{figure}
\mbox{}\hfill\epsfysize1.25\hsize\epsfbox{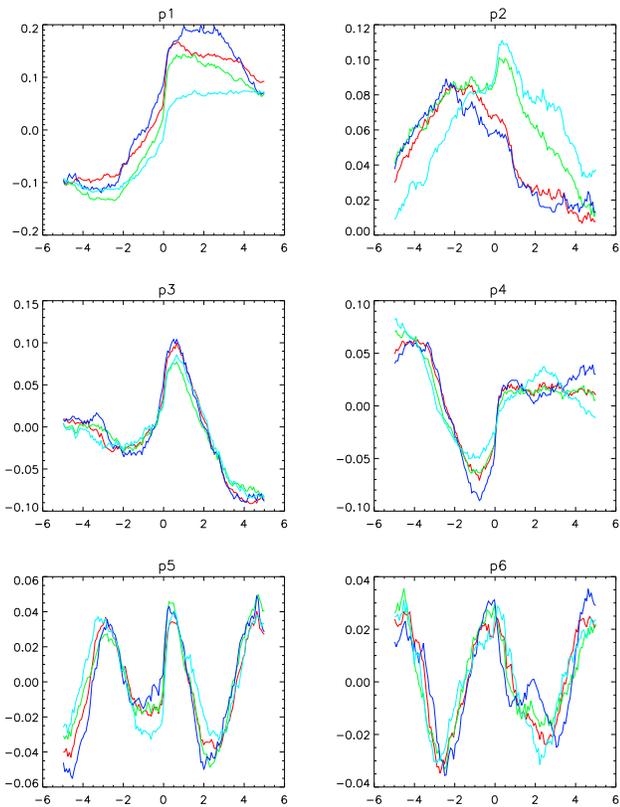}\hfill\mbox{}
\caption{\label{halves}Stability of the pincipal components with respect to time variability of the
data. The full set of correlation functions has been divided into halves in two ways. The dark
lines show the even and odd numbered samples (chronologically ordered), measuring
differences on a time scale of 25 s. The lighter lines show the first and second halves of the set,
indicating differences on a time scale of hours to days. The higher components appear to be 
quite stable in this sense, but in fact most of these turn out to be artefacts (see text and Figs.
\ref{synthout}).
}
\end{figure}

The nature of the remaining components is not obvious at this point. Some of them, such as $P_3$
and $P_5$, appear to contain signals related to the sharp rise of the cross correlation at $t=0$.
The PCA also interprets pure noise in the data in terms of additional components, and it is not a
priori obvious which components are systematic and which may be just artefacts. A useful test to
see which of the components is stable with respect to noise is to divide the set of samples in two
halves and comparing the components derived from each of these subsets. This is shown in Fig.\
\ref{halves}. The four curves shown result from two different ways of dividing the set in halves.
In the first, the two subsets consist of the even numbered and the odd numbered samples of the
(chronologically ordered) set. The difference between these two subsets measures the intrinsic
variability of the cross correlation on a time scale of 25s (the duration of a sample). In the second
division, the first half of the set is compared with the second half. This measures the variability
on time scales of days. 

\begin{figure}[t]
\mbox{}\epsfxsize1.0\hsize\epsfbox{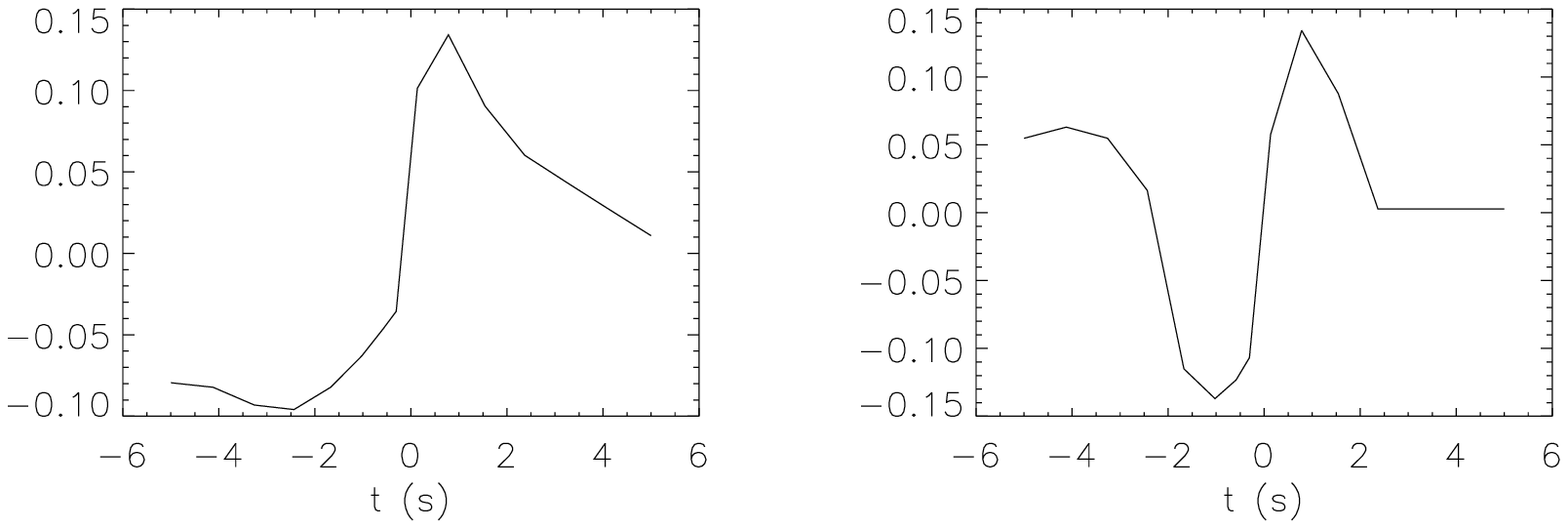}\hfill\mbox{}
\mbox{}\epsfysize0.35\hsize\epsfbox{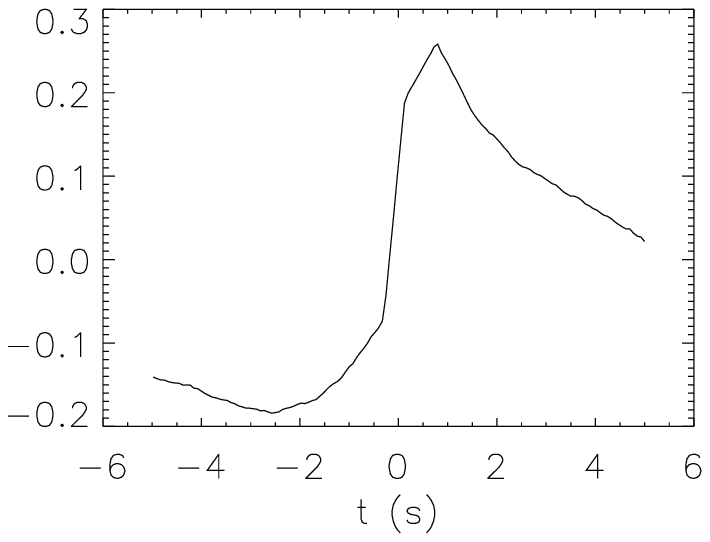}\hfill\mbox{}
\caption{\label{synthin}Top: input components for the synthetic set of cross correlations. Their 
amplitudes are varied independently. In addition there is a noise component (not shown) whose 
contribution to the variance peaks around 0.2 Hz.  The average cross correlation produced by
these components is shown in the bottom panel.
}
\end{figure}

As Fig.\ \ref{halves} shows, the components, especially the higher ones, are remarkably stable
on both the short and the long time scales. From this one would be tempted to conclude that all
six components shown are real. The process by which the components are derived in PCA,
however, leaves the possibility open that some of the components are artefacts. This is because
the components are constructed as orthogonal directions in the abstract $n-$dimensional space of
variables. The directions of insignificant higher components (constructed out of noise) are thus
biased in a systematic way by the directions of the significant lower components.

The only way to test which of the components is real in this sense is by comparison of the results
with a synthetic data set.

\subsubsection{Tests with synthetic data}
To test the significance of the components, {we have constructed synthetic data and applied
PCA on this data in the same way as was done with the real  data. By varying the input data,
the resulting PCs can be adjusted for agreement. Since many of the components found are
suspected to be spurious, the aim of the excercise was to find the minimum number of
independently varying signals needed to  reproduce the components found by PCA of the real
data. It is not neccessary to start the synthetic process with time series of the count rates, since
the questions to be answered concern the behavior of principal component analysis as applied to
our cross-correlation functions. Thus, the synthetic data constructed consist of samples of the
cross correlation function. Each of the samples consists of a linear combination of a number of
adjustable functions of the time delay coordinate $t$, here called `signals'. The shape of these
signals is the same in all samples, but their amplitudes differ. In addition to these signals, we 
add a random noise component that is different in each sample.

The amplitudes of the signals are assigned} randomly between the samples, normally distributed
with specified means and deviations. The noise component consists of white noise, low-pass
filtered by boxcar averaging with a width $\delta t$. The parameters of the synthetic data are
thus: the shapes of the signals, the means $A_{\rm 1}$ and $A_{\rm 2}$ and deviations
$\sigma_{\rm 1}$,  $\sigma_{\rm 2}$ of their amplitudes, the amplitude $A_{\rm n}$ of the noise
component, and $\delta t$.

\begin{figure}[hb]
\mbox{}\hfill\epsfysize1.0\hsize\epsfbox{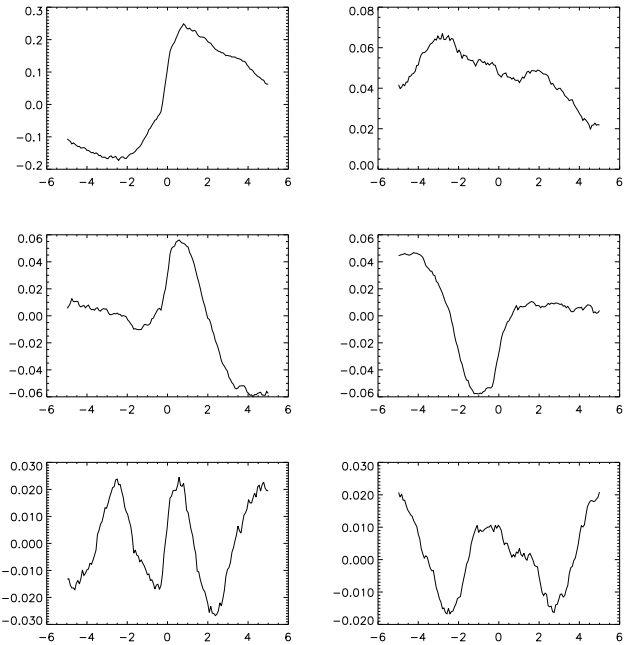}\hfill\mbox{}
\caption{\label{synthout} Components found by PCA of a sample of 297 synthetic cross
correlations, constructed from the two input components shown in Fig.\ref{synthin}. The shapes
of  the input components and their amplitudes were adjusted for agreement of the resulting
synthetic PCA components with the first 6 components of the actual data (Fig. \ref{allsix})
}
\end{figure}

PCA on a set of 297 of these synthetic cross correlations yields the components shown in Fig.
\ref{synthout}, {for a case with two input signals plus a noise component. The shapes of the
two input signals were adjusted to fit the shapes of the output components to those of the real
data (Fig. \ref{allsix}). The final result for these input signals is shown in fig. \ref{synthin}

The amplitude of first two PCs of the simulated data are related directly to the average
amplitudes of the two input signals. The amplitudes of PCs 3 and higher, however, turn out to be
proportional to the amplitude of the boxcar-filtered noise component instead, showing that they
are artefacts of the PCA. PCA somehow constructs these higher components out of the noise
component. The {\it shapes} of these higher components, however, depend 
sensitively on the shape of the two input signals. The input signals can be determined rather
accurately by the fitting process. Two signals plus a noise component turned out to be the
minimum number needed. Tests with only one signal plus noise have also been done.} These are
able to reproduce the first two PCA components, but leave most of the detail in the higher
components unexplained.

Except for the first, none of the components found by the PCA resembles the  input 
{signals} very well. It is thus clear that PCA without comparison with a simulation using
synthetic data can not be used to identify the statistically independent components reliably in the 
present problem. Unfortunately, the forward modeling process was quite cumbersome, owing to 
the large number of parameters to be adjusted, and is not easily automated.

From this test we conclude that just two independent signal components are needed to explain
the observed variability of the cross correlation, apart from an aditional unsystematic noise
component. {The interpretation of these components} presents some puzzles. One might have
hoped that the actual independent signals would be the `spike' and the `precognition dip', but
this turned out not to be the case. Extensive attempts were made to reproduce the PCA
components with input {signals} resembling a pure dip before $t=0$ and a spike after $t=0$,
but no satisfactory fit
was found. When the shape of the {signals} was relaxed away from this prejudice, {
iteration converged onto the signals shown} in Fig.\ \ref{synthin}. Their shapes could be
determined rather accurately by this fitting process. 

Both {signal components} consist of a dip followed by a peak. The second {signal
component} resembles a compressed (in time) version of the first, but with the peak being
somewhat less pronounced relative to the dip. A possible interpretation is thus that the intrinsic
shape of the optical response is a combination of a dip followed by a sharp peak, while the time
scale of this pattern varies, by a factor of 3 or so. 

\begin{figure}
\mbox{}\hfill\epsfxsize1.0\hsize\epsfbox{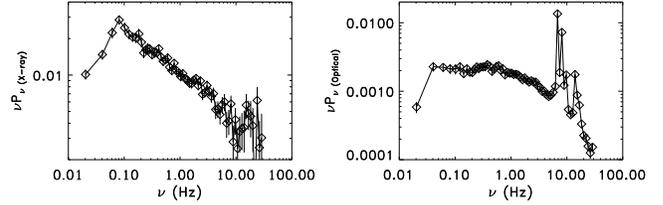}\hfill\mbox{}
\caption{\label{psp} Average power spectra of the X-ray and optical time series (unit on y-axis:
variance per factor $e$ in frequency). The peaks in the optical spectrum around 10 Hz are
artefacts due to telescope vibrations. 
}
\end{figure}

\begin{figure}
\mbox{}\hfill\epsfxsize1.0\hsize\epsfbox{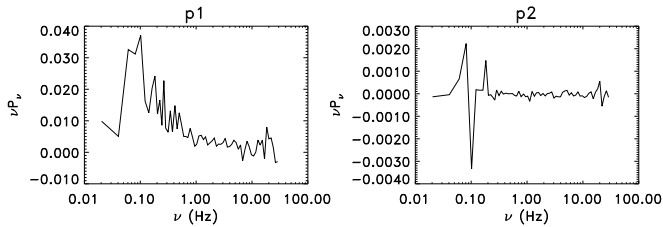}\hfill\mbox{}
\caption{\label{pcsp} Principal components of the variability in the X-ray power spectra. (Note 
that the vertical scale is linear, instead of the log scale in Fig.\ \ref{psp})
}
\end{figure}

\section{Power spectra}
Information complementary to the cross correlations is given by the power spectra. The average
X-ray and optical power spectra are shown in Fig. \ref{psp}. The broad peak near 0.1 Hz
in the X-ray power spectrum is associated with a QPO around this frequency. This QPO has been
found in most of the X-ray data from KV UMa (Revnivtsev et al. 2000, Wood et al. 2000), and
with high  amplitude also in the optical light curve (Haswell et al. 2000, Pavlenko et al. 2001).  In
our data its amplitude was 
significantly smaller than in these earlier observations, and undetectable in the optical. Its 
frequency varied a bit during our observations, with some preference for frequencies of 0.08 and
0.2 Hz. 

The variation of the power spectrum can be studied with dynamical power spectra; such studies
have 
been reported for many sources (cf. van der Klis 2000). In the spirit of our analysis of the cross
correlation, the variability in the power spectra can also by studied by PCA. Treating the 297
individual power spectra in the same way as the cross correlations yields components shown in
Fig. \ref{pcsp}.

The first component of the  X-ray power spectrum is rather similar to the average, with a broad
peak around 0.1 Hz. The second component shows a dip at 0.1 Hz and peaks at 0.08 Hz and 0.2
Hz. This could be explained {by} the presence of QPOs at frequencies of 0.08, 0.1 and 0.2
Hz,
with the amplitude of the 0.1 Hz peak decreasing when the others are increasing. Tests with
synthetic data show, however, that PCA also produces such a pattern if there is just a {\it single}
a peak at a somewhat variable frequency. This seems a more likely interpretation in view of what
is known about QPOs in similar sources.

The optical power spectra are affected by telescope vibrations at frequencies near 10 and 40 Hz.
As explained above, these vibrations were caused by the rather strong wind loading of the
telescope on several of the nights. The optical variability is also contaminated with seeing effects,
which have frequencies in the range range of interest here. Since the X-O cross correlation
reaches maximum values around 40\% it is unlikely that seeing effects dominate the power
spectrum, but its precise shape can not be reliably deduced from our observations. 

With this caveat, the optical power spectrum shown in Fig. \ref{psp}  is much broader than the
X-ray power spectrum, and there is no clear evidence for the QPO seen in the X-rays. It thus
seems unlikely that a QPO contributes strongly to the X-O cross correlations. The effect of the
X-ray QPO's on our cross correlations may well be just a a random noise contribution. 

\section{Conclusions and discussion}
We have studied the correlation of X-ray and optical properties of the black hole transient KV
UMa. Thanks to the length of the dataset, the statistical properties of this correlation could be
studied in detail. By analysis of the time series themselves, we find that the optical response to an
X-ray peak consists of a broad dip followed by a sharper peak. By cross-correlating the two time
series the same pattern is found, but with a much higher signal-to-noise ratio.  From the cross
correlation we find that the onset of the peaks in the optical response coincides with the X-ray
maximum to within 30 ms, and their initial rise is very fast. The minimum of the dip in the optical
precedes the X-rays by about 2 seconds.  

The agreement of the pattern of the cross-correlation with that found directly in the time series 
establishes, on the one hand, that the dip and the delayed peak are both properties of the
optical light curve rather than the X-rays, and on the other hand that they are not due to
artefacts of background subtraction or the cross-correlation process. 

The shape of the cross correlation was found to be highly variable on time scales from minutes to
days. A principal component analysis (PCA) of this variablity was found to be of some value in
studying its statistical properties. It turned out that the principal components themselves, except
for the first, are not very useful since their shape is highly affected by artefacts arising from a
strong, randomly variable additive noise component in the cross correlation. 

We find that the PCA components can be used, however, as an intermediate representation of
the data, which can then be interpreted by a forward modeling process. The result of such a
simulation shows that there are two systematic components which vary statistically
independently. Both have the shape of a broad dip followed by a sharper peak, but their
duration differs by a factor of about 3.

The most puzzling aspect of the optical variations is of course the `precognition dips'. If the
optical light is due to reprocessing of X-rays, then an optical signal preceding the X-rays would
have to be due to something in the accretion flow preceding an X-ray event. In principle, this
could happen if the increase in mass flux which {somewhat later} is observed as an increase
{of} the X-ray flux, starts at some distance from
the hole, such that the accretion time scale from that distance is of the order of a few seconds.
For typical accretion disk models, this would be a distance intermediate between the X-ray
emitting region and the optical reprocessing regions in the outer disk. If this event is associated
with a local thickening of the accretion disk,  and the X-ray emitting region is small enough, such 
a `bump' on the disk surface might be sufficient to shield the outer disk temporarily from the
X-rays, leading to a dip in the optical emission. 

While this possibility may be attractive in the reprocessing model, reprocessing itself has
difficulties explaining the shape of the optical response. The sudden change in slope at zero lag
turns out to be impossible to reproduce with any positive definite disk reprocessing function. The
main reason for this is that the sudden change in slope requires much more variability on very
short time scales ($<100$ ms) than the X-ray light curve actually has. A second difficulty is that
the optical-to-X-ray flux ratio in KV UMa was quite high, or the order 10\%. If, as the spectral
energy distribution of McClintock et al. suggests, the UV-EUV emission is part of the same
process producing the visible light, the reprocessed flux would have to be as large as the X-rays
themselves. 

From these difficulties it appears that alternatives may have to be explored to explain the visible
light in black hole accreters like KV UMa and GX 339-4. In a future paper we will explore the
possibility that the visible light is produced as thermal cyclo-synchrotron emission in the inner
20000 km of the accretion disk, as first suggested by Fabian et al. (1982) for GX 339-4 (see 
also Merloni et al. 2001 for an interpretation of the spectrum of KV UMa). If this
interpretation is correct, it suggests a new and potentially very powerful means of probing the
inner accretion flow by means of optical, and especiallly coordinated X-ray and optical
observations.

\acknowledgements{We thank the XTE time allocation team and Skinakas observatory for their
support and flexibility in making these observations possible, Dr.\ C.\ Straubmeier and F.\ Schrey, who carried out the optical observations and Dr.\ T.\ Belloni, who provided the XTE count rate data. We thank Dr. Christian Motch for his comments on an earlier version of this
text. This work was done with support from the European Commission under grant no
ERB-FMRX-CT98-0195.}  

{}

\end{document}